\documentclass[twocolumn,prb,showkeys,showpacs,preprintnumbers,amsmath,amssymb]{revtex4}

\usepackage{amssymb}
\usepackage{graphicx}
\usepackage{dcolumn}
\usepackage{amsmath}
\usepackage{bm}
\usepackage{epsfig}

\begin{document}
\title{Path integral representation of the evolution operator for the Dirac equation}
\author{Alexander S. Lukyanenko}
\email{alex.lukyan@rambler.ru}
\affiliation{Department of Experimental Physics, St. Petersburg State
Polytechnical University, Polytekhnicheskaya 29, 195251, St. Petersburg,
Russia }
\author{Inna A. Lukyanenko}
\email{inna.lukyan@mail.ru}
\affiliation{Department of Mathematics, St.Petersburg State University,
Universitetski prospect 28, Stary Petergof, 198504, St. Petersburg, Russia}

\begin{abstract}
A path integral representation of the evolution operator for the
four-dimensional Dirac equation is proposed. A quadratic form of the
canonical momenta regularizes the original representation of the path
integral in the electron phase space. This regularization allows to obtain a
representation of the path integral over trajectories in the configuration
space, i.e. in the Minkowsky space. This form of the path integral is useful
for the formulation of perturbation theory in an external electromagnetic
field.
\end{abstract}

%\pacs{}
\maketitle

\date{\today}

%\begin{multicols}{2}
%\narrowtext

%%%%%%%%%%%%%%%%%%%%%%%%%%%%%%%%%%%%%%%%%%%%%%%%%%%%%%%%%%%%%%%%%%%%%%% 

The formulation of quantum mechanics in terms of complex amplitudes and its
concentrated representation in terms of the Feynman path integral in a
configuration space became very useful for the further development of
quantum theory. In particular, this approach is the most suitable for the
formulation of perturbation theory rules. Moreover, it is difficult to
formulate the quantum theory of gauge fields without the Feynman path
integral. However, one of the oldest parts of quantum theory, the
relativistic quantum mechanics of an electron, which is based on the Dirac
equation, has no path integral representation. Feynman has attempted to
solve the problem \cite{f}. He hoped to understand the nature of the spin by
use of this representation. He obtained a solution for an electron in one
dimension (1D). In this case two components of an electron wave function are
related to two possible ways to reach the end point of a path. For an
electron in three dimensions (3D) such a clear solution was not obtained. In
this paper we return to the problem of a path integral representation of
solutions of the Dirac equation. We consider this representation as a tool
for the formulation of perturbation theory rules for an evolution operator
of the Dirac equation in an external electromagnetic field.

The Dirac equation for an electron in an external electromagnetic field
described by the four-vector potential $A_{\mu }=\left( \varphi
,A_{i}\right) $ has a form \cite{bd}: 
\begin{eqnarray}
i\hbar \frac{\partial \psi }{\partial t} &=&\widehat{h}\psi \equiv c\widehat{%
\alpha }_{k}\left( \frac{\hbar }{i}\nabla _{k}-\frac{e}{c}A_{k}\right) \psi +
\nonumber \\
&&+\widehat{\beta }mc^{2}\psi +e\varphi \psi  \label{1}
\end{eqnarray}
Here $\psi \ $is the Dirac bi-spinor, and the matrices $\widehat{\alpha }%
_{k}\ $and $\widehat{\beta }\ $are defined by the following commutational
relations: 
\begin{eqnarray}
\widehat{\alpha }_{i}\widehat{\alpha }_{k}+\widehat{\alpha }_{k}\widehat{%
\alpha }_{i} &=&2\delta _{ik},\widehat{\alpha }_{i}\widehat{\beta }+\widehat{%
\beta }\widehat{\alpha }_{i}=0,  \nonumber \\
\widehat{\beta }^{2} &=&1,i,k=1,2,3  \label{2}
\end{eqnarray}
Let us introduce an evolution operator for the Dirac equation (\ref{1}) in a
time interval $\left[ t^{\prime },t^{\prime \prime }\right] $:

\begin{equation}
\widehat{U}\left( t^{\prime \prime },t^{\prime }\right) \equiv T\exp \left( -%
\frac{i}{\hbar }%
%TCIMACRO{\dint}%
%BeginExpansion
\displaystyle\int %
%EndExpansion
\limits_{t^{\prime }}^{t^{\prime \prime }}\widehat{h}dt\right)  \label{3}
\end{equation}
Here, $T$ denotes the time ordering of the exponent. The problem is how to
represent the kernel of this operator, $\left\langle x^{\prime \prime
}\left\vert \widehat{U}\right\vert x^{\prime }\right\rangle \equiv \widehat{K%
}\left( x^{\prime \prime },t^{\prime \prime };x^{\prime },t^{\prime }\right) 
$ , as an integral over trajectories in the 3D configuration space of an
electron. Let us notice that this kernel is a $4\times 4$- matrix in the
Dirac bi-spinor space. Below this kernel will be called \textquotedblleft
propagator\textquotedblright .

Following the standard prescription \cite{fs}, let us begin with the
definition of the integral over trajectories in the phase space. Here, we
meet the following problem. In the non-relativistic mechanics the
Hamiltonian $\widehat{h}\left( p,x\right) $ is quadratic with respect to the
canonical momenta. As a consequence the transition to the path integral in
the configuration space is not difficult. Contrary to this case, the
Hamiltonian of the Dirac equation is a linear matrix operator with respect
to the canonical momenta. A formal integration over the canonical momenta in
this case leads to a meaningless expression $\delta ^{3}\left( \stackrel{.}{x%
}^{i}-c\widehat{\alpha }^{i}\right) $. In order to overcome this difficulty,
we will regularize the path integral in the phase space, adding
\textquotedblleft by hands\textquotedblright\ a quadratic form of the
canonical momenta. In account of this, let us start with the building of the
Feynman path integral.

Let us divide the time interval $\left[ t^{\prime },t^{\prime \prime }\right]
$\ into $N$\ parts $\Delta t_{k}\equiv \left[ t_{k},t_{k-1}\right] $\ of
equal length $\varepsilon =\left( t^{\prime \prime }-t^{\prime }\right) /N$
, where $k=1,...,N$\ , $t_{0}\equiv t^{\prime },t_{N}\equiv t^{\prime \prime
}$\ . Basic trajectories in the phase space are broken lines in the $3D$%
-configuration space with pieces $\Delta x_{\left( k\right) }^{i}\equiv
x_{\left( k\right) }^{i}-x_{\left( k-1\right) }^{i}$, where we define $%
x_{\left( 0\right) }^{i}\equiv \left( x^{i}\right) ^{\prime },x_{\left(
N\right) }^{i}\equiv \left( x^{i}\right) ^{\prime \prime }$. The momentum in
any time interval $\Delta t_{k}$\ has a constant value $p_{i\left( k\right)
} $.

The standard procedure of iterations leads to the following representation
for the propagator of the Dirac equation (1) in terms of an integral in the
phase space:

\begin{eqnarray}
\widehat{K} &=&%
%TCIMACRO{
%\underset{%
%\begin{array}{c}
%N\rightarrow \infty \\ 
%\sigma \rightarrow 0%
%\end{array}
%}{\lim }}%
%BeginExpansion
\mathrel{\mathop{\lim }\limits_{%
\begin{array}{c}
N\rightarrow \infty \\ 
\sigma \rightarrow 0%
\end{array}
}}%
%EndExpansion
\int \frac{d^{3}p_{\left( N\right) }}{\left( 2\pi \hbar \right) ^{3}}%
%TCIMACRO{\dprod}%
%BeginExpansion
\mathop{\displaystyle\prod}%
%EndExpansion
\limits_{k=1}^{N-1}\times  \nonumber \\
&&\times \frac{d^{3}x_{\left( k\right) }d^{3}p_{\left( k\right) }}{\left(
2\pi \hbar \right) ^{3}}\times  \nonumber \\
&&\times \exp \left[ -\frac{\sigma p_{\left( k\right) }^{2}}{2}+\frac{i}{%
\hbar }p_{i\left( k\right) }\Delta x_{\left( k\right) }^{i}\right] \times 
\nonumber \\
&&\times T%
%TCIMACRO{\dprod}%
%BeginExpansion
\mathop{\displaystyle\prod}%
%EndExpansion
\limits_{k=1}^{N}\left[ 1-\frac{i}{\hbar }\varepsilon \widehat{h}\left(
p_{\left( k\right) },\widetilde{x}_{\left( k\right) }\right) \right]
\label{4}
\end{eqnarray}
Here $\widetilde{x}_{\left( k\right) }^{i}\equiv \left( x_{\left( k\right)
}^{i}-x_{\left( k-1\right) }^{i}\right) /2$, and

\begin{eqnarray}
\widehat{h}\left( p_{\left( k\right) },\widetilde{x}_{\left( k\right)
}\right) &\equiv &c\widehat{\alpha }_{i}\left( p_{i\left( k\right) }-\frac{e%
}{c}A_{i}\left( \widetilde{x}_{\left( k\right) }\right) \right) +  \nonumber
\\
&&+\widehat{\beta }mc^{2}+e\varphi \left( \widetilde{x}_{\left( k\right)
}\right)  \label{5}
\end{eqnarray}
The latest multiplier in the Eq.(\ref{4}) is the time ordered product.

In account of the regularizing quadratic form, the integration over the
canonical momenta may be easily performed. As a result, we obtain an
integral over trajectories in the configuration space:

\begin{eqnarray}
\widehat{K} &=&%
%TCIMACRO{
%\underset{%
%\begin{array}{c}
%N\rightarrow \infty \\ 
%\sigma \rightarrow 0%
%\end{array}
%}{\lim }}%
%BeginExpansion
\mathrel{\mathop{\lim }\limits_{%
\begin{array}{c}
N\rightarrow \infty \\ 
\sigma \rightarrow 0%
\end{array}
}}%
%EndExpansion
\frac{1}{\left( 2\pi \hbar \sigma \varepsilon \right) ^{3/2}}\int 
%TCIMACRO{\dprod}%
%BeginExpansion
\mathop{\displaystyle\prod}%
%EndExpansion
\limits_{k=1}^{N-1}\frac{d^{3}x_{\left( k\right) }}{\left( 2\pi \hbar \sigma
\varepsilon \right) ^{3/2}}\times  \nonumber \\
&&\times \exp \left( -\frac{\Delta x_{\left( k\right) }^{2}}{2\hbar
^{2}\sigma \varepsilon }\right) \times  \nonumber \\
&&\times T%
%TCIMACRO{\dprod}%
%BeginExpansion
\mathop{\displaystyle\prod}%
%EndExpansion
\limits_{k=1}^{N}\left[ 1-\frac{i}{\hbar }\varepsilon \widehat{h}\left( 
\frac{i}{\hbar }\frac{\Delta x_{\left( k\right) }}{\sigma \varepsilon },%
\widetilde{x}_{\left( k\right) }\right) \right]  \label{6}
\end{eqnarray}
This representation of the propagator is convenient for the formulation of
the perturbation theory.

First of all, let us demonstrate that in the zero order of the electron
charge the integral in Eq.(\ref{6}) is equal to the well known propagator of
a free electron. To this end, let us introduce a momentum representation of
the propagator: 
\begin{eqnarray}
K\left( \Delta t,p^{^{\prime \prime }},p^{^{\prime }}\right) &=&\int
d^{3}x^{\prime \prime }\int d^{3}x^{\prime }\widehat{K}\left( x^{\prime
\prime },t^{\prime \prime };x^{\prime },t^{\prime }\right) \times  \nonumber
\\
&&\times \exp \left( -\frac{i}{\hbar }p^{\prime \prime }x^{\prime \prime }+%
\frac{i}{\hbar }p^{\prime }x^{\prime }\right)  \label{7}
\end{eqnarray}
The integration over coordinates of the end points of a trajectory in Eq.(%
\ref{7}) allows to perform the following transformation of the integration
variables in Eq.(\ref{6}): $x_{\left( k\right) }^{i}\rightarrow \Delta
x_{\left( k\right) }^{i},i=1,...,N$ , where we take into account that

\begin{equation}
\left( x^{i}\right) ^{\prime \prime }=\left( x^{i}\right) ^{\prime }+%
%TCIMACRO{\dsum}%
%BeginExpansion
\mathop{\displaystyle\sum}%
%EndExpansion
\limits_{k=1}^{N}\Delta x_{\left( k\right) }^{i}  \label{8}
\end{equation}
After this, the integral in Eq.(\ref{6}) may be easily calculated. The
integration over $\left( x^{i}\right) ^{\prime }$\ gives a multiplier $%
\left( 2\pi \hbar \right) ^{3}\delta ^{3}\left( p^{\prime \prime }-p^{\prime
}\right) $. The integral over $\Delta x_{\left( k\right) }^{i}$\ is equal to

\begin{eqnarray}
&&%
%TCIMACRO{\dprod}%
%BeginExpansion
\mathop{\displaystyle\prod}%
%EndExpansion
\limits_{k=1}^{N}\hbar ^{3}\left( 2\pi \sigma \varepsilon \right) ^{3/2}\exp
\left( -\frac{\sigma \varepsilon \left( p^{\prime \prime }\right) ^{2}}{2}%
\right) \times  \nonumber \\
&&\times \left[ 1-\frac{i}{\hbar }\widehat{h}\left( p^{\prime \prime
}\right) \right]  \label{9}
\end{eqnarray}
Here,

\begin{equation}
\widehat{h}\left( p^{\prime \prime }\right) \equiv c\widehat{\alpha }%
_{k}p_{\left( k\right) }^{\prime \prime }+\widehat{\beta }mc^{2}  \label{10}
\end{equation}
is the Hamiltonian of a free electron. The time order of the matrix
multiplies in Eq.(\ref{9}) is unessential. Combining these results, one can
obtain the propagator of a free electron:

\begin{eqnarray}
\widehat{K}^{\left( 0\right) }\left( \Delta t;p^{\prime \prime },p^{\prime
}\right) &=&\left( 2\pi \hbar \right) ^{3}\delta ^{3}\left( p^{\prime \prime
}-p^{\prime }\right) \times  \nonumber \\
&&\times \exp \left( -\frac{i}{\hbar }\widehat{h}\left( p^{\prime \prime
}\right) \Delta t\right)  \label{11}
\end{eqnarray}

A contribution of the first order in the electric charge to the propagator
is equal to

\begin{eqnarray}
\widehat{K}^{\left( 1\right) }\left( t^{\prime \prime },t^{\prime
};p^{\prime \prime },p^{\prime }\right) &=&ie%
%TCIMACRO{\dint}%
%BeginExpansion
\displaystyle\int %
%EndExpansion
\limits_{t^{\prime }}^{t^{\prime \prime }}dt\exp \left( -\frac{i}{\hbar }%
\widehat{h}\left( p^{\prime \prime }\right) \left( t^{\prime \prime
}-t\right) \right) \times  \nonumber \\
&&\times \widehat{A}\left( t,p^{\prime \prime }-p^{\prime }\right) \times 
\nonumber \\
&&\times \exp \left( -\frac{i}{\hbar }\widehat{h}\left( p^{\prime }\right)
\left( t-t^{\prime }\right) \right)  \label{12}
\end{eqnarray}
Here

\begin{equation}
\widehat{A}\left( t,q\right) \equiv \int d^{3}x\exp \left( -\frac{i}{\hbar }%
qx\right) \left[ \widehat{\alpha }_{k}A_{k}\left( t,x\right) -\varphi \left(
t,x\right) \right]  \label{13}
\end{equation}
is the momentum representation of an external electromagnetic field
contribution.

In conclusion, using a regularizing quadratic form of the electron momentum
in the phase space, we have obtained a representation of the one-electron
relativistic quantum dynamics in terms of a path integral over trajectories
in the $3D$-configuration space. This representation may be used as a basis
for the formulation of the relativistic quantum dynamics of many-electron
systems with the direct interaction.

%%%\noindent $^{\ast }$ E-mail address: alex.lukyan@rambler.ru

%%%\noindent $^{+}$ E-mail address: inna.lukyan@mail.ru

%\begin{references}

%%\newpage

\end{document}